# Spin and Valley Polarized Multiple Fermi Surfaces of α-RuCl$_3$/Bilayer Graphene Heterostructure

Soyun Kim[1†], Jeonghoon Hong[2,3†], Kenji Watanabe[4], Takashi Taniguchi[5], Joseph Falson[6], Jeongwoo Kim[2*], and Youngwook Kim[1*]

[1]*Department of Physics and Chemistry, Daegu Gyeongbuk Institute of Science and Technology (DGIST), Daegu 42988, Republic of Korea*

[2]*Department of Physics, Incheon National University, Incheon 22012, Republic of Korea*

[3] *Department of Physics, Indiana University, Bloomington, IN 47405, USA*

[4]*Research Center for Electronic and Optical Materials, National Institute for Materials Science, Tsukuba 305-0044, Japan*

[5]*International Center for Materials Nanoarchitectonics, National Institute for Materials Science, Tsukuba 305-0044, Japan*

[6]*Department of Applied Physics and Materials Science, California Institute of Technology, Pasadena, CA 91125, USA*

[†]These authors contributed equally

Authors to whom correspondence should be addressed: kjwlou@inu.ac.kr and y.kim@dgist.ac.kr

**Abstract**

We report the transport properties of α-RuCl$_3$/bilayer graphene heterostructures, where carrier doping is induced by a work function difference, resulting in distinct electron and hole populations in α-RuCl$_3$ and bilayer graphene, respectively. Through a comprehensive analysis of multi-channel transport signatures, including Hall measurements and quantum oscillation, we unveil significant band modifications within the system. In particular, we observe the emergence of spin and valley polarized multiple hole-type Fermi pockets, originating from the spin-selective band hybridization between α-RuCl$_3$ and bilayer graphene, breaking the spin degree of freedom. Unlike α-RuCl$_3$/monolayer graphene system, the presence of different hybridization strengths between α-RuCl$_3$ and the top and bottom graphene layers leads to an asymmetric behavior of the two layers, confirmed by effective mass experiments, resulting in the manifestation of valley-polarized Fermi pockets. These compelling findings establish α-RuCl$_3$ proximitized to bilayer graphene as an outstanding platform for engineering its unique low-energy band structure.

The assembly of different materials has emerged as a promising strategy for modifying the low-energy band structure through the proximity effect.[1-3] Creating heterostructures has been a well-established approach in the thin-film epitaxy community. However, the combination of arbitrary materials has been significantly limited due to considerations of lattice mismatch between the materials and substrates. In contrast, within the van der Waals community, the stacking of different 2D materials allows for desired modifications to the band structure without geometric constraints.[4-7] For example, by bringing graphene into proximity with materials possessing a large spin-orbit coupling, such as $WSe_2$, the spin-orbit coupling of graphene can be significantly enhanced.[8-14] Moreover, twisting has emerged as a highly effective technique that utilizes interlayer coupling between similar materials to alter the low-energy band structure, create a flat band near the Fermi level [15-22], and induces a reconstructed lattice in twisted $WTe_2$ and $α-MoO_3$.[23-25]

In particular, exploring the proximity effect between magnetic and non-magnetic van der Waals materials presents fascinating possibilities for band engineering. It enables the opening of a bandgap in bilayer graphene when in proximity to magnetic 2D materials like chromium trihalides.[26] A interfacial interaction between graphene and CrOCl gives rise to a SiC-graphene type quantum Hall phase up to 80 K.[27,28] Similar non-monotonic gate dependences have also been observed in the graphene/$CrI_3$,[29] further demonstrating the potential for band engineering and interfacial effects. Furthermore, the magnetic proximity effect allows for the hybridization of spin-polarized materials with non-magnetic materials, leading to the emergence of a spin-split band structure within the non-magnetic materials.[30,31] However, this intriguing property in the vicinity of the Fermi level is exclusively observed in specific material combinations, as it is significantly affected by the difference in work function between them. As a result, capturing a spin-selective conduction channel remains a challenging task.

$α-RuCl_3$/graphene heterostructure is a promising candidate for exploring the magnetic proximity effect in 2D materials. Spin-polarized bands originating from the zig-zag type antiferromagnetic $α-RuCl_3$ are hybridized with the Dirac bands of graphene at the Fermi level, giving rise a spin-split Fermi surface.[32-37] The quantum oscillations with extraordinary temperature dependence and thermal signals suggest the presence of spin fractionalization and the potential existence of Majorana fermions in $α-RuCl_3$.[38-40] However, despite the fascinating quantum phenomena of α-RuCl$_3$, the exploration of hybridization between α-RuCl$_3$ and other two-dimensional materials of high quality is still lacking, which could offer further possibilities for exploitable band engineering and strong magnetic proximity coupling. Furthermore, while spin splitting can be controlled through proximity coupling, manipulating the valley degree of freedom remains a significant challenge. The valley degree of freedom can only be manipulated in graphene when it is aligned with the hBN structure.[5-7,41] The detection of a valley-polarized Fermi pocket has not yet been achieved, presenting an ongoing hurdle in this field of research.

In order to identify the potential of band engineering and investigate the proximity effect, we study the transport properties of α-RuCl$_3$/bilayer graphene heterostructures. Firstly, we have observed that the work function difference between α-RuCl$_3$ and bilayer graphene leads to heavy hole-doping in the bilayer graphene. Additionally, the interaction between antiferromagnetic band of α-RuCl$_3$ and bilayer graphene induces spin-selective band hybridization, breaking the spin degree of freedom within the bilayer graphene system. In contrast to previous studies on α-RuCl$_3$/monolayer graphene, our analysis of effective mass has confirmed a strong hybridization between the α-RuCl$_3$ band and the top layer of graphene, while the hybridization with the bottom layer is comparatively weaker. This interplay between the layer and valley indices in bilayer graphene results in the lifting of the valley degree of freedom, allowing us to identify the presence of spin-polarized and valley-polarized Fermi pockets within the system. Additionally, we explored the effective mass in higher energy bands of bilayer graphene, which had remained unexplored until now. Moreover, we discovered that the band structure of α-RuCl3/bilayer graphene exhibits a highly sensitive response to an applied electric field, resulting in a phenomenon known as magnetic breakdown. This phenomenon, specific to bilayer systems, represents a truly unique aspect when compared to monolayer systems.

α-RuCl$_3$ differs from conventional 2D materials, as it is sensitive to acetone exposure, making traditional lithography and metallization methods unlikely to succeed.[34,42] To prevent the degradation of α-RuCl$_3$, we modified the fabrication procedure by first preparing the bottom electrodes and subsequently transferring the heterostructure onto them (The details are available in the Methods section of Supplementary Information and other relevant references [34,42]). Here we focus on our highest quality device, although data from other devices is available in the Supplementary Information. Figure 1a shows $R_{xx}$ plotted against the magnetic field at a $V_g = 0$ V and $T = 2$ K. The quantum oscillations exhibit two characteristics. First, we found rapid oscillations that exhibit a beating mode which is attributed to two or more comparable oscillation frequencies. The second feature is an oscillation with a slower frequency. These findings demonstrate the emergence of at least two similar-sized Fermi surfaces and a smaller Fermi surface than the former pockets. Besides multiple quantum oscillations, the Hall resistance exhibits a nonlinear curve in the inset of Fig.1b, suggesting that multiple charge carriers are involved in the transport measurements.[34]

To perform an accurate analysis, we subtracted the background oscillations and plotted $\Delta R_{xx}$ as a function of $1/B$ in Fig. 1c. We noticed two changes with adjusting the backgate voltage from 60 V to -60 V: (i) the number of nodes for slow oscillations increased, and (ii) the intensity of fast oscillations decreased. FFT analysis facilitated a simpler comparison, as shown in Fig. 1d. In the low-frequency range, two dominant peaks are observed below 100 T, for example, at 16 T and 30 T for $V_g = 60$ V. These peaks gradually shift towards higher frequencies until reaching 65 T and 81 T at $V_g = -60$ V, indicating two hole-type Fermi pockets. While extra peaks were observe, such as 12 T and 24 T at $V_g = 0$, they do not consistently depend on gate voltage in left panel of Fig. 1c, which may be attributed to low-frequency background noise. The high-frequency peaks exhibit a similar trend to the low-frequency region, but have four peaks: two prominent peaks (blue and green triangles) and two minor peaks (purple and yellow triangles). These peaks shifted to higher frequencies as the gate voltage decreased. Additional peaks were discovered in the high-frequency range, as marked by gray triangles, but they appeared when the back gate voltage was negative, and the resulting oscillations were very weak. In contrast to the six peaks depicted in Fig. 1e, these particular peaks do not demonstrate any discernible systematic dependence on the gate voltage.Therefore, it was challenging to determine the origin of the three bluish FFT outputs shown in Fig. 1d. However, we noticed significant FFT peaks at 60 V, which were associated with strong oscillations and could not be ignored.

To elucidate the origin of the FFT peaks, we investigated the electronic structure of the α-RuCl$_3$/bilayer graphene heterostructure as shown in Fig. 2a. We reproduced the observed charge transfer from graphene to α-RuCl$_3$. The hole-doped bilayer graphene states were observed at the K point, exhibiting weak hybridization with the flat bands originating from α-RuCl₃. In contrast to α-RuCl₃/monolayer graphene, the proximity of the α-RuCl₃ layer to bilayer graphene results in the breaking of valley symmetry due to the strong interaction with the top graphene layer. This leads to distinct band dispersions along the K-M and K'-M lines as exhibited in Fig. 2b. Notably, Figure 2c and 2d demonstrates that the hybridization between the top graphene layer (Gra1) and the adjacent α-RuCl₃ layer (gray) induces band splitting along the K-M line (α region) while leaving the K'-M' line unaffected (δ region). The hybridized bands naturally become spin-polarized due to the magnetic order of the α-RuCl₃ layer, as depicted in Fig. 2d and S2c. This anisotropic graphene band induced by the hybridization with the α-RuCl₃ suggests the presence of different Fermi wavevectors, potentially correlated with the multiple peaks in the high-frequency region. Furthermore, the presence of two distinct low-frequency peaks can be attributed to the spin splitting at the K or K' point, as illustrated in Fig. 2d and Fig. S2b. Although the bottom layer (Gra2) is not in direct contact with the α-RuCl₃ layer, the graphene bands at the K point primarily originating from it are spin-polarized. The broken valley symmetry may lead to the emergence of one or two additional peaks in the low-frequency region, contingent upon the Fermi level. However, the magnitude of this splitting is marginal, posing challenges in its experimental identification.

We computed the Fermi surface of the α-RuCl₃/bilayer graphene heterostructure and investigated its variation with the binding energy. As exhibited in Fig. 2e, the α-RuCl₃ possesses a lemon-shaped electron pocket (red line) centered at the Γ point, while the bilayer graphene exhibits inner (small $K_f$) and outer (large $K_f$) hole pockets (blue line) located at the corners of the Brillouin zone (K point) (Fig. 2f). For pure bilayer graphene, there are spin-degenerate Fermi pockets at the K point. However, due to the hybridization with α-RuCl₃, one of the outer hole pockets from the top graphene layer undergoes slight distortion, resulting in the formation of a smaller, closely packed Fermi surface. This can be observed as the red pocket at the K point in Fig. 2e (indicated by red arrow). The emergence of spin-dependent red pocket at K indicates the existence of two distinct oscillations in the high-frequency region. In addition, the broken valley symmetry of K and K' induces a slight variation in the size of the outer pockets depending on the valley (Fig. S2c), thereby leading to the occurrence of additional oscillation peaks, as observed in Fig. 1. Figure 2g shows that the Fermi surface driven by the bilayer graphene (blue line) is well maintained against hole/electron doping whereas the α-RuCl₃ states (red line) are highly sensitive to the Fermi energy. In the negative binding energies (or hole doping), two concentric hole pockets monotonically increase in size while the electron pocket shrinks to the Γ point, which is consistent with the trends shown in Fig. 1. In the positive binding energies, the separated electron pockets become connected at the M point and start to form an open Fermi surface and the hole pocket coming from the bottom graphene layer (Gra2) disappears. At 10 meV, the Fermi pockets at the K point and the open Fermi surface at Γ point are in close proximity, leading to possibility of the formation of different carrier orbit paths. This phenomenon, known as magnetic breakdown, is particularly likely to occur and results in the appearance of additional peaks in the high-frequency region at $V_g$ = 60 V, as illustrated in Fig. 1d.

We analyze the band mass of the inner and outer Fermi surfaces to identify the valley polarization in the α-RuCl$_3$/bilayer graphene heterostructure. The band mass quantifies the hybridization strength between the α-RuCl$_3$ layer and each graphene layer. Specifically, the α-RuCl$_3$ band component leads to an increase in the effective mass of the bilayer graphene band, making it significantly heavier. To determine the cyclotron, we performed temperature-dependent quantum oscillations and subtracted the fast and slow background $R_{xx}$, respectively. We plotted the isolated Shubnikov-de Haas oscillations (at $V_g$ = 0 V) for each fast and slow case in Fig. 3a and b, after subtracting the respective backgrounds. The temperature dependence of the singular peak, denoted by a star, in each panel is presented in Fig. 3a and 3b. (See section 4 of Supplementary Information for details)

To evaluate the effective masses, we have fitted the data points with temperature factor ($R_T$) of Lifshitz-Kosevich formula:

$$R_T = \frac{\alpha p \mu T/B}{\sinh(\alpha p \mu T/B)}$$

with $\alpha = 2\pi^2 m_e k_B/eh \approx 14.69\ (T/K)$, where $p$ is harmonics, $h$ is the Planck constant, $m_e$ is the free electron mass, $k_B$ is the Boltzmann constant, $e$ is the elementary charge. The $\mu$ represents the ratio of cyclotron mass to the free electron mass, $m^*/m_e$, where $m^*$ denotes the cyclotron mass. We have found that the value of $m^*/m_e$ for fast oscillations (Fig. 3c) is 0.07056, whereas for slow oscillations (Fig. 3d), it is 0.04523. This discrepancy suggests a notable difference in band hybridization between the inner/outer Fermi surfaces and the α-RuCl$_3$ Fermi surface. These results align with our theoretical calculations, as depicted in Fig. 2c and 2e. Note that the slow oscillation was exclusively observed using the ionic liquid gate technique applied to bilayer graphene. However, there has been no previous analysis of band mass in this specific context. The α-RuCl$_3$/bilayer graphene heterostructure presents a unique advantage that underscores its distinctive characteristics

In conclusion, we demonstrate the magnetic proximity effect in the α-RuCl$_3$/bilayer graphene heterostructure through various transport measurements, including Hall measurements and Shubnikov-de Haas oscillations. The interaction between α-RuCl$_3$ and the top layer of bilayer graphene leads to spin-selective band hybridization, resulting in the breaking of the spin degree of freedom in bilayer graphene. Interestingly, we discovered an intrinsic valley-polarized Fermi pocket, which arises from the stronger hybridization between the top layer of graphene and the α-RuCl$_3$ band. This observation was further confirmed through effective mass analysis and theoretical calculations, revealing the distinct properties of our heterostructure that sharply contrast with those of the α-RuCl$_3$/monolayer graphene heterostructure. Overall, our findings shed light on the complex interplay between α-RuCl$_3$ and bilayer graphene, encompassing band modifications, and spin and valley degrees of freedom.

See the supplementary material for the additional magnetotransport data, extra effective mass data, theoretical calculation for breaking of the spin and valley symmetry in bilayer graphene, six Fermi pockets, lithium intercalation data, theoretical calculation for electronic structure of α-RuCl$_3$/bilayer graphene heterostructure with Li intercalation, and experimental and theoretical methods.

**Acknowledgment**

This work was supported by the Basic Science Research Program NRF-2020R1C1C1006914, the DGIST R&D program (23-CoE-NT-01) of the Korean Ministry of Science and ICT, the DGIST-Caltech collaboration research program (23-KUJoint-01), and Samsung electronics. K.W. and T.T. acknowledge support from the JSPS KAKENHI (Grant Numbers 20H00354, 21H05233 and 23H02052) and World Premier International Research Center Initiative (WPI), MEXT, Japan. J.K. was supported by the National Research Foundation of Korea funded by the Korea government (MSIT) (No. NRF-2022R1F1A1059616)

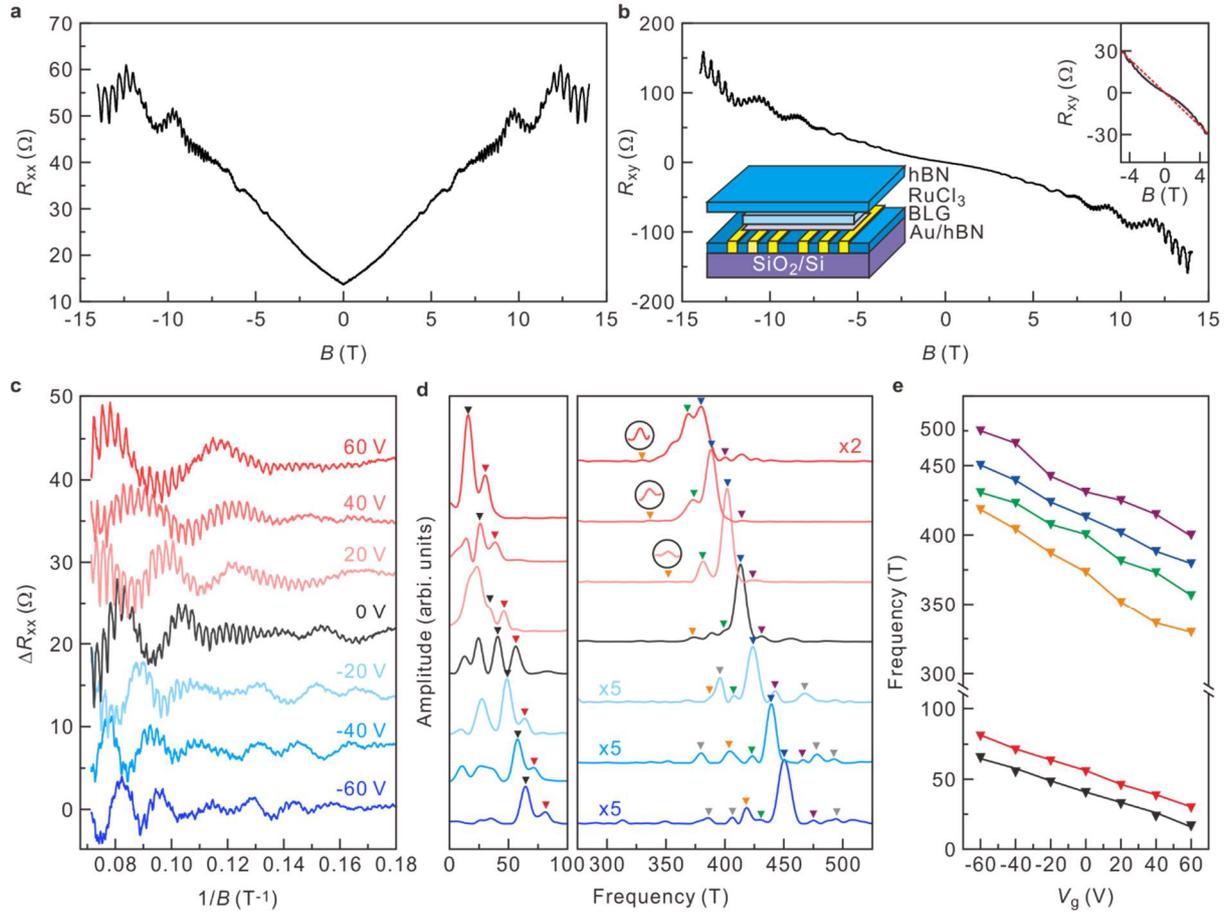

**Figure 1 a,** Longitudinal resistance as a function of the magnetic field at $T = 2$ K and $V_g = 0$ V for bilayer graphene/RuCl$_3$. **b,** Same as **a** but for the Hall resistivity. Inset displays a magnified view of the original $R_{xy}$ data, highlighting the non-linear $R_{xy}$ curve. The red dotted line represents the linear fitting curve of $R_{xy}$. **c,** $\Delta R_{xx}$ as a function of the inverse magnetic field. Different colors indicate different gate voltages at $T = 2$ K. The inset shows schematic of device. **d,** FFT spectra of Shubnikov-de Haas oscillations of **c**. Triangles mark six frequency peaks. The inserted graphs with circles represent magnified data of yellow triangle-marked peaks multiplied ten times for each dataset. **e,** Gate voltage-dependent peak sets represented in **d**. The color codes are the same as the color of the triangles in **d**.

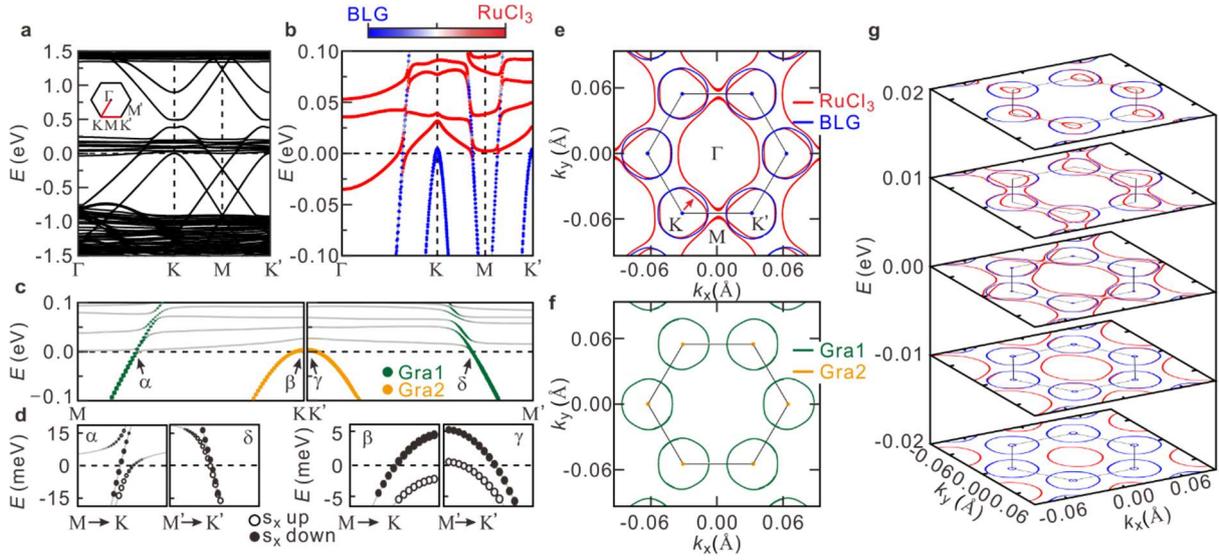

**Figure 2 a,** Calculated band structure of α-RuCl$_3$/bilayer graphene heterostructure along the high symmetry line, as indicated in the inset. The Fermi level set to zero. **b,** Atomically projected band structure of the heterosrturcture near the Fermi level, with blue and red dots representing the atomic contribution from bilayer graphene (BLG) and α-RuCl$_3$, respectively. **c,** Calculated band structures for the heterostructure, shown with graphene layer resolution (upper). Green and orange dots represent the top layer graphene adjacent to α-RuCl$_3$ (Gra1) and the bottom layer graphene (Gra2), respectively. **d,** Same as **c** but with spin resolution near the Fermi level. Empty and filled dots denote the spin-up and spin-down components, respectively. **e,** Calculated Fermi surface with atomic resolution, where blue and red lines represent BLG and α-RuCl$_3$, respectively. Red arrow indicates distorted band at K-point. **f,** Calculated Fermi surface with atomic resolution, where green and orange lines represent the top graphene layer and the bottom graphene layer, respectively. **f,** Variation in the Fermi surface in an energy range between -0.02 and +0.02 eV. blue and red lines represent BLG and α-RuCl$_3$, respectively.

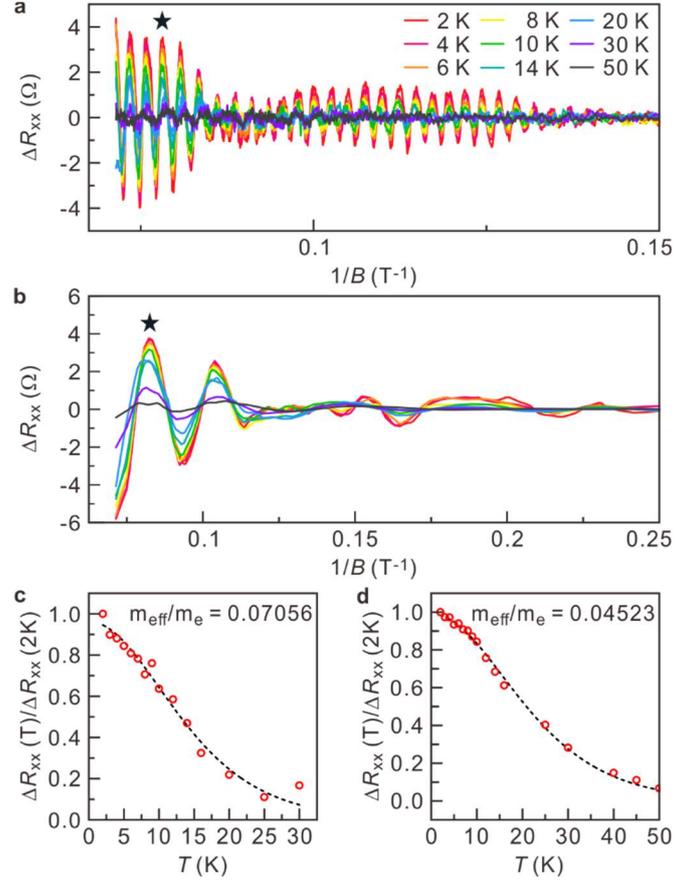

**Figure 3 a**, Fast oscillations isolated from original $\Delta R_{xx}$ as a function of inverse magnetic field at $V_g = 0$ V. Different colors indicate different temperatures from 2 K to 50 K. **b**, Same with **a** but for slow oscillation. **c**, $\Delta R_{xx}$ normalized by its $R_{xx}$ at $T = 2$ K and $1/B = 0.078$ T$^{-1}$ as a function of temperature for the fast oscillations. Dotted line is fitting curve with Lifshitz- Kosevich formula. **d**, Same with **c**, but for slow oscillations at $1/B = 0.082$ T$^{-1}$.

Supplementary Information

# Spin and Valley Polarized Multiple Fermi Surfaces of α-RuCl$_3$/Bilayer Graphene Heterostructure

Soyun Kim[1†], Jeonghoon Hong[2,3†], Kenji Watanabe[4], Takashi Taniguchi[5], Joseph Falson[6], Jeongwoo Kim[2*], and Youngwook Kim[1*]

[1]*Department of Physics and Chemistry, Daegu Gyeongbuk Institute of Science and Technology (DGIST), Daegu 42988, Republic of Korea*

[2]*Department of Physics, Incheon National University, Incheon 22012, Republic of Korea*

[3] *Department of Physics, Indiana University, Bloomington, IN 47405, USA*

[4]*Research Center for Electronic and Optical Materials, National Institute for Materials Science, Tsukuba 305-0044, Japan*

[5]*International Center for Materials Nanoarchitectonics, National Institute for Materials Science, Tsukuba 305-0044, Japan*

[6]*Department of Applied Physics and Materials Science, California Institute of Technology, Pasadena, CA 91125, USA*

[†]These authors contributed equally
[*]E-mail: kjwlou@inu.ac.kr and y.kim@dgist.ac.kr

**S1. Additional magnetoransport data**

As mentioned in the main manuscript, we made modifications to the fabrication procedure. Initially, we prepared the bottom electrodes (Fig. S1a and b) and then transferred the heterostructure onto them, as illustrated in Fig. S1c and d. The atomic force microscopy image in Fig. S1d was captured after conducting magnetotransport measurements. In Fig. S1c and 3d, second device consists of two distinct regions. The left five electrodes are dedicated to the bilayer graphene, while the right set is connected to the α-RuCl$_3$/bilayer graphene heterostructure.

Magnetotransport data for second device were obtained at $V_g$ = -40 V and $T$ = 2 K for bilayer graphene and a α-RuCl$_3$/bilayer graphene heterostructure, as illustrated in Fig. S1e and f, respectively. We observed two distinct trends, despite both originating from the same bilayer graphene. First, the resistance in the absence of a magnetic field exhibited different characteristics: the bilayer graphene component displayed a resistance of 50 Ω, whereas the heterostructure only showed a resistance of 16 Ω. Second, there was a notable contrast in the frequency of Shubnikov-de Haas oscillations between the bilayer graphene and heterostructure regions. The former exhibited slower oscillations, while the latter displayed faster ones. To compare the frequencies more clearly, we conducted a fast Fourier transform (FFT) analysis as depicted in Fig. S1g. The FFT peak of the bilayer graphene section (highlighted in blue) was detected at ~20 T, and the peak at around 400 T (indicated in purple) corresponded to the heterostructure. Here, the frequency can be converted to density using $f = hn/ge$, where $h$ is the Planck constant, $n$ is carrier density, $g$ is the degeneracy factor, and $e$ is the elementary charge. The 20 T peak corresponds to the typical charge density of a bilayer graphene device at $V_g$ = -40 V, which is approximately $1.9 \times 10^{12}$ cm$^{-2}$ with $g$ = 4 (spin and valley). In Fig. S1g, we observed a relatively broad single peak centered around 400 T, which suggests preserving internal spin and valley degrees of freedom. This is in contrast to the behavior observed in our main data, where a spin and valley-polarized Fermi surface was detected. The absence of this expected feature could potentially be attributed to disorder-induced potential fluctuations arising from bubbles, thus masking the detection of the spin-polarized Fermi surface.

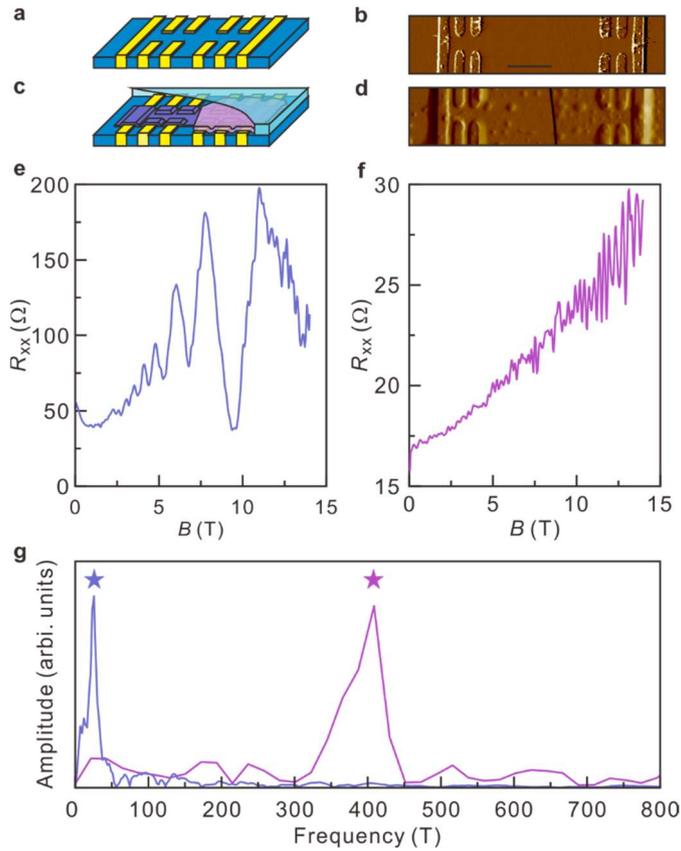

**Figure S1 a**, Schematic of integrated Au electrodes with bottom hBN. **b**, Atomic force microscopic image of **a**. Scale bar is 5 μm. **c,** Same as **a** but with bilayer graphene/RuCl$_3$/top hBN. The bilayer graphene with blue color covers entire electrodes while five electrodes on the right side are connected to bilayer graphene/RuCl$_3$ (purple color) heterostructure. Both areas are encapsulated with hBN layers. **d,** Atomic force microscopic image of **c**. The bold line in the center shows the boundary of RuCl3. **e,** Longitudinal resistance as a function of the magnetic field at $T$ = 2 K and $V_g$ = -40 V for the bilayer graphene region. **f,** Same as **e** but for bilayer graphene/RuCl$_3$ area. **g,** FFT spectra of the **e** and **f**. Each color code matches up with the specified area in illustration **c**.

## S2. Breaking of the spin and valley symmetry in bilayer graphene

The graphene bands near the Fermi level exhibit spin polarization, as shown in Figure S2. The presence of the α-RuCl$_3$ layer breaks time-reversal symmetry in the heterostructure, leading to a distinct separation between up and down spin states. Furthermore, the presence of the α-RuCl$_3$ layer causes a lifting of the valley degeneracy in the bilayer graphene, resulting in the inequivalence of the two graphene layers. Consequently, the band dispersions and Fermi wavevectors vary depending on the **k** path.

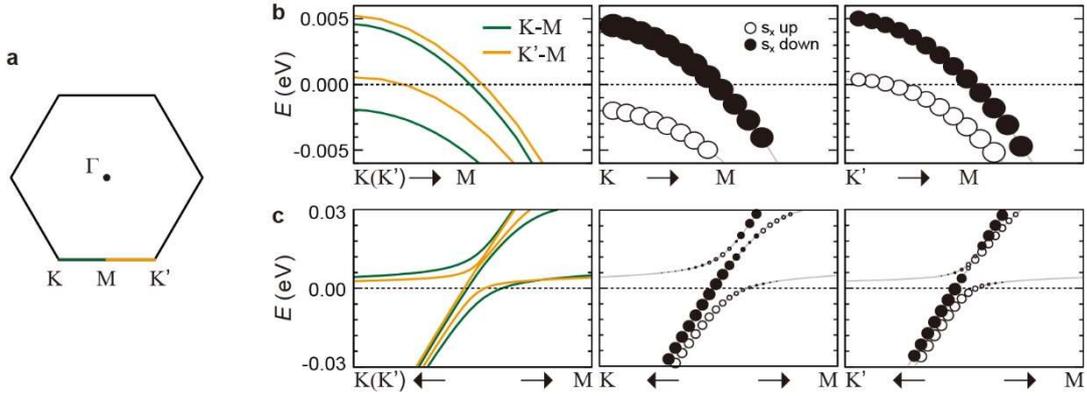

**Figure S2 a,** The first Brillouin zone of α-RuCl$_3$/ bilayer graphene heterostructure. **b,** (Left panel) Calculated band structure at near K and K' points. The band dispersions along the K-M line (green) and K'-M line (orange) are overlaid. (Middle panel) Calculated band structure with spin resolution along the K-M line. (Right panel) Calculated band structure with spin resolution along the K'-M line. Empty and filled dots denote the spin-up and spin-down components, respectively. **c,** (Left panel) Calculated band structure in the middle of the K-M (green) and K'-M (orange) lines. (Middle panel) Calculated band structure with spin resolution along the K-M line. (Right panel) Calculated band structure with spin resolution along the K'-M line. Red and blue dots denote the spin-up and spin-down components, respectively.

## S3. Multiple Fermi pockets

We assign the six frequencies in the Fig S3 to their respective Fermi surfaces based on the results presented in Figure 1c.

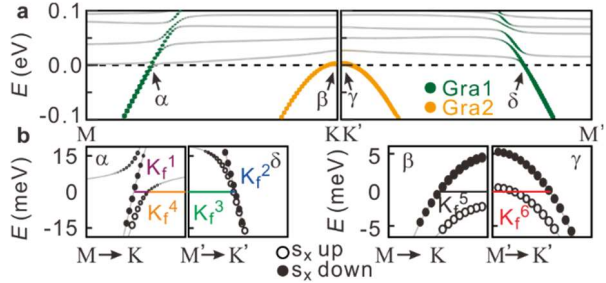

**Figure S3 a, b,** Same as **Fig 2c and d** but with six K$_f$.

## S4. Extra effective mass data

Figure S4a illustrates the procedure used to determine the background of fast/slow oscillations. Initially, the maxima and minima values of the original data were extracted and interpolated. The median value of the two sets was represented by a solid blue line, indicating the background of fast oscillations. A distinct slow oscillation was also represented by a blue line. Subsequently, the same process was repeated, with the solid blue line as the initial target. The resulting background of slow oscillations was represented by solid red lines. During this process,

In the main manuscript, we focus on the analysis of a single peak to determine the effective mass, which represents the most fitting result among the various peaks observed. However, in order to provide a more comprehensive understanding, we also present additional data in Figure S4, which includes a wider range of peaks. All results from fast and slow oscillations consistently show similar outputs.

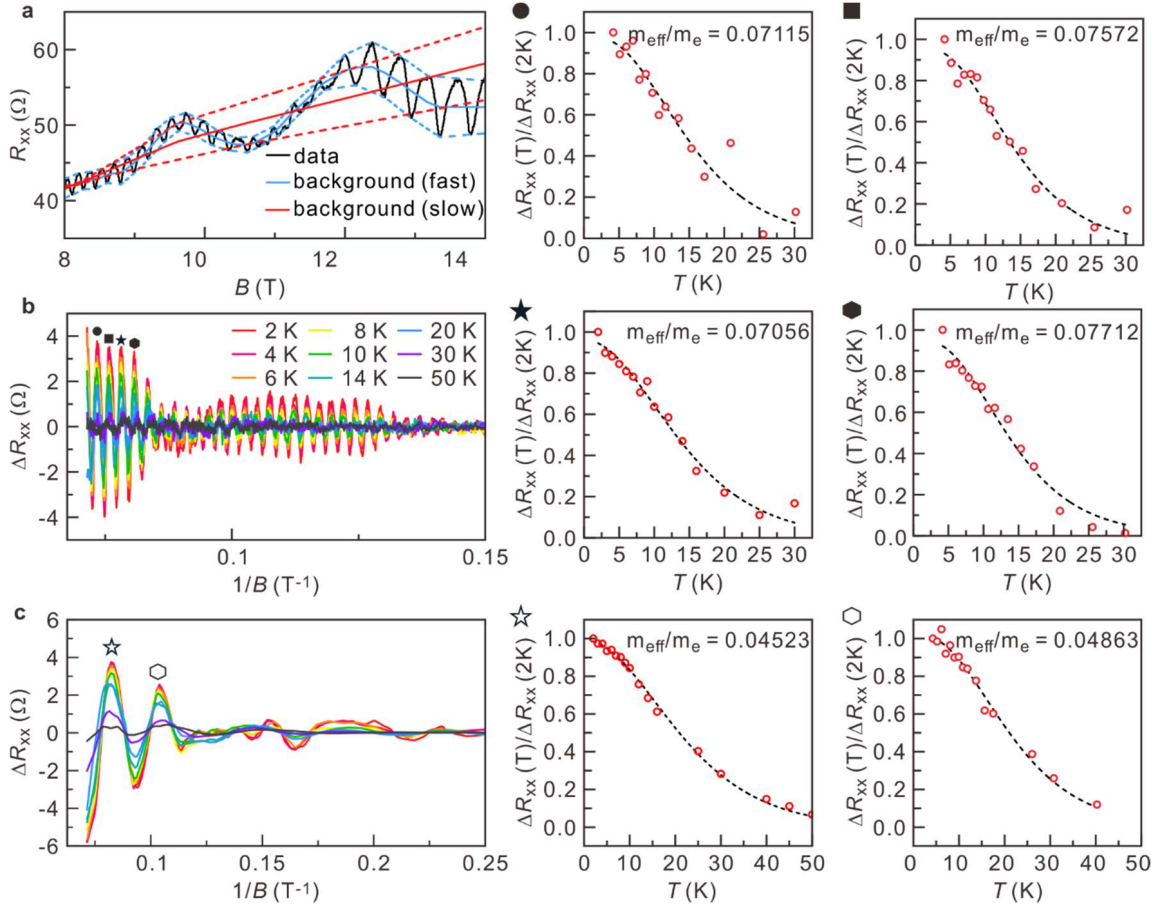

**Figure S4 a**, Shubnikov-de Haas oscillation at $T = 2$ K and $V_g = 0$ V. Dashed blue lines are originated from the interpolation of $R_{xx}$ oscillation extrema. Blue solid line represents background of fast oscillations obtained from average of two blue dashed lines. Blue solid line itself implies isolated slow oscillations. Red dashed and solid lines are similar with blue dashed/solid lines, but for slow oscillations. The red solid line represents background of slow oscillations and integrated oscillations. **b**, **c** $\Delta R_{xx}$ as a function of inverse magnetic field for fast and slow oscillations with various temperature, respectively. These figures correspond to the same plots presented in the main manuscript as Fig 3.b and c. The symbols in the figures represent the oscillation nodes used for the mass analysis, incorporating the Lifshitz-Kosevich formula.

## S5. Lithium intercalation to α-RuCl$_3$/bilayer graphene

With the lithium intercalation to heterostructure, it is possible to alter the Fermi pockets from a mixed electron/hole type to electron Fermi pockets, theoretically characterized by an electron density of $1.63 \times 10^{14}$cm$^{-2}$, as depicted in Fig. S5a. While conventional liquid gating techniques are commonly employed to modulate high-density charge carriers, it is important to note that these methods are unsuitable due to the reactivity of the liquid with α-RuCl$_3$. In light of this, we have elaborated on the advantages of employing lithium intercalation in our heterostructure as an alternative means to achieve the desired band structure. To this end, we use a lithium-ion conducting glass-ceramic substrate with a potential gradient within our heterostructure.[S1-S5] This gradient causes lithium ions to migrate towards the bilayer graphene. Figure S5b illustrates the longitudinal resistance as a function of the backgate voltage. We bias the gate voltage to zero volts and cool down the device to a temperature of $T = 2$ K. Subsequently, we measure the $R_{xy}$ for this state. The orange curve in Fig. S5c represents the $R_{xy}$ curve obtained for $V_g = 0$ V, exhibiting a non-linear behavior similar to the inset of Fig. 1b, indicating multi-channel conduction. Next, we warm up the device to 300 K and increase the voltage to 4.6 V at a rate of 10 mV/sec. During this process, we observed resistance peaks, which serve as charge neutrality points. These peaks suggest that lithium donates electrons, shifting the chemical potential of the bilayer graphene towards a positive direction. After cooling the device back down to 2 K while maintaining $V_g = 4.6$ V, we repeat the $R_{xy}$ measurement. The $R_{xy}$ curve of the lithium-intercalated heterostructure continues to exhibit a linear relationship with the electron carrier density with $n = 1.4 \times 10^{14}$cm$^{-2}$, as demonstrated by the navy curve in Fig. S5b. This doping level aligns with findings from our theoretical prediction and previous studies on Li-intercalated bilayer graphene.[S1,S2] Finally, to revert to a hole-doped bilayer graphene, we proceed to remove the lithium ions by gradually decreasing the gate voltage. As a result, the lithium ions migrate back to the substrate. As expected, this process results in the observation of a resistance peak again, indicating the return to a hole-doped state in the bilayer graphene. We observed hysteresis, which is most likely attributed to the relatively fast gate sweep rate. This phenomenon is commonly observed in ionic gate experiments and can account for the observed hysteresis in our measurements.

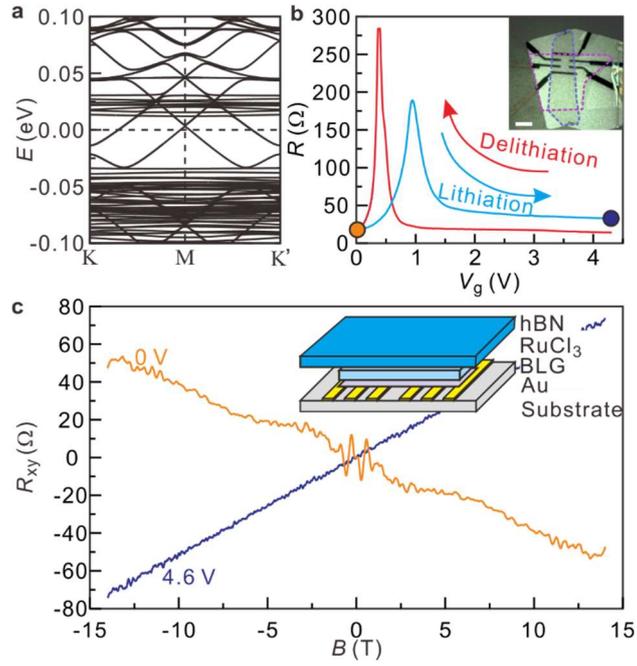

**Figure S5 a**, Band structure of a Li-intercalated α-RuCl$_3$/bilayer graphene heterostructure. **b**, Backgate dependence of longitudinal resistance at room temperature. The lithiation process is indicated by the blue curve, showing an increase in voltage bias, while the delithiation process is represented by the red curve. The orange and navy dots correspond to the 0V and 4.6V bias points for Hall measurements without and with lithium in the heterostructure, respectively. The inset displays the heterostructure on top of a lithium-ion conducting glass-ceramic substrate, with a 10 μm scale bar. **c**, symmetrized Hall curves as a function of perpendicular magnetic field at $T = 2$ K. Different colored curves represent different gate voltage conditions. The inset shows schematic of device for lithium intercalation.

## S6. Electronic structure of α-RuCl₃/bilayer graphene heterostructure with Li intercalation

We investigated the effect of Li interaction on the electronic structure of the heterostructure as shown in Figure S6. We considered two different Li configurations: between the graphene layers, and between the bottom graphene layer and the α-RuCl₃ layer. Whereas the Li atoms adjacent to the α-RuCl₃ layer marginally shift the Fermi level, as exhibited in Fig. S6c, the Li intercalation between the bilayer graphene gives rise to a highly *n*-doped electronic state, as shown in Fig. S6d.

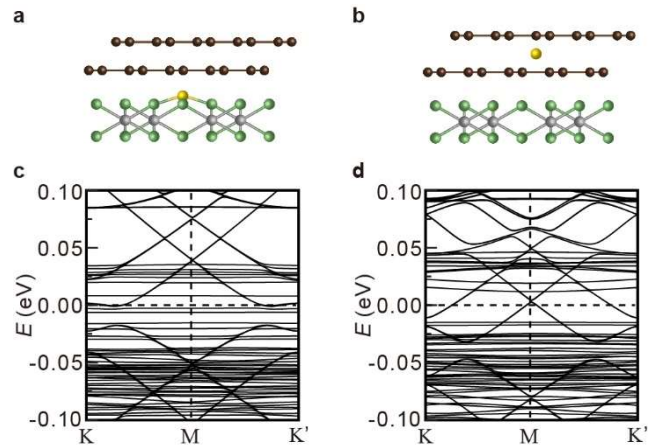

**Figure S6** The atomic structure of α-RuCl₃/bilayer graphene heterostructure with an intercalated Li atom: **a** between the bilayer graphene layers, and **b** between the α-RuCl3 and the bottom graphene layers. The brown, green, gray and yellow spheres represent C, Cl, Ru and Li atoms, respectively. The corresponding calculated band structures for both configurations are shown in **c** and **d**. The localized spins of the Ru atoms are deliberately aligned in a ferromagnetic manner in this calculation.

## S7. Methods

### Device Fabrication

We initially searched for pristine hBN flakes at around 30 nm. Next, using E-beam lithography, we fabricated multiple electrodes directly on top of the hBN flake. After developing PMMA, we etched the electrode area with a reactive ion etching machine using a gas mixture of $CF_4$ (40 sccm) and $O_2$ (4 sccm) at a power of 40 W. We then filled the etched area with Cr/Au. The thickness of the metals is 0.5~1 nm higher than hBN to ensure contact between the heterostructure and the bottom electrodes, as shown in Fig. 1a. We immerse the sample in warm acetone for approximately 12 hours during the lift-off process to minimize PMMA residue on the surface of the hBN region. Any remaining PMMA on the left is removed through mechanical wiping using the contact mode of atomic force microscopy, as shown in Fig. 1b. The bottom electrode substrates are now prepared. We employ the standard pick-up transfer technique to create a heterostructure using a motorized x-y-z stage. First, we pick up the top hBN layer using an elvacite stamp. The identical procedure is replicated for the combination of α-RuCl$_3$ and graphene. Next, we transfer the entire heterostructure onto the bottom electrode, as depicted in Fig. 1c. In principle, the α-RuCl$_3$ within our device is unresponsive to acetone due to the full encapsulation of the device by top and bottom hBN layers. However, we typically do not melt leftover elvacite resin.

### Device for Lithium Intercalation

Our device structure is similar to the device described in conventional device. However, in our setup, there is no bottom hBN substrate. Instead, in order to intercalate lithium to bilayer graphene, the bilayer graphene directly rests on a lithium-ion conducting glass-ceramic substrate. This modification enables the intercalation of lithium into the bilayer graphene directly, specifically through the edges of the bilayer.

### Computational Method

Our density functional theory computations were performed employing the projected augmented plane-wave method [S6,S7], as implemented in the Vienna Ab Initio Simulation Package (VASP) [S8]. The Perdew-Burke-Ernzerhof (PBE) type generalized gradient approximation (GGA) was utilized to describe the exchange-correlation interactions among the electrons. We considered the van der Waals (rev-vdW-DF2) [S9,S10] corrections to describe the weak interaction between BLG and α-RuCl$_3$. The energy convergence threshold for the self-consistent field iteration was set at $10^{-5}$ eV. The energy cutoff for the plane-wave basis expansion is chosen to be 400 eV. A (5 × 5) supercell of graphene atop a ($\sqrt{3} \times \sqrt{3}$) α-RuCl$_3$ layer with a 15 Å vacuum along the surface normal direction was utilized to simulate α-RuCl$_3$/BLG heterostructure. A dense 15 × 15 × 1 **k**-point grid was employed to sample the Brillouin zone. The lattice parameter of the slab structure was fixed to the graphene layer and the atomic positions were relaxed until the total energy reached a threshold of less than $10^{-3}$ eV. To describe the electronic correlation effect of *d* orbitals, we invoke the Hubbard-*U* correction (GGA+*U* method) [S11,S12] for Ru *d* orbitals ($U$ = 2 eV). To account for the Fermi level pinning in the heterostructure, we applied an artificial electric field of -75 meV/Å along the z-axis. This was necessary to reproduce the two distinct graphene Fermi surfaces observed in our experiment. We utilized the WANNIER90 code to construct a tight-binding Hamiltonian and obtained the Fermi surface of the heterostructure using 1000 × 1000 × 1 **k** points.

**Reference for Supplementary Information**